\documentclass[twocolumn, aps, pra,prl, superscriptaddress, preprintnumbers, letterpaper]{revtex4-1}

\usepackage{amsthm}
\usepackage{amsmath,bm}
\usepackage{amssymb}
\usepackage{braket}
\usepackage{bbold}
\usepackage{graphicx}
\usepackage{subfigure}
\usepackage{hyperref}
\usepackage{appendix}
\usepackage{tikz}
\usepgflibrary{shapes.geometric}
\usetikzlibrary{arrows}

\hypersetup{colorlinks=true, citecolor=blue, urlcolor=blue, linkcolor=blue}

\def\la{\langle}

\begin{document}
\title{Operational Entanglement Detection Based on $\Lambda$-Moments}
\author{Ke-Ke Wang}
\email{wangkk@cnu.edu.cn}
\affiliation{School of Mathematical Sciences, Capital Normal University, 100048 Beijing, China}
\author{Zhi-Wei Wei}
\email{weizhw@cnu.edu.cn}
\affiliation{School of Mathematical Sciences, Capital Normal University, 100048 Beijing, China}
\author{Shao-Ming Fei}
\email{feishm@cnu.edu.cn (Corresponding author)}
\affiliation{School of Mathematical Sciences, Capital Normal University, 100048 Beijing, China}
\affiliation{Max-Planck-Institute for Mathematics in the Sciences, 04103 Leipzig, Germany}

\bigskip

\begin{abstract}
We introduce $\Lambda$-moments with respect to any positive map $\Lambda$. We show that these $\Lambda$-moments can effectively characterize the entanglement of unknown quantum states without theirs prior reconstructions. Based on $\Lambda$-moments necessary and sufficient separability criteria, as well as necessary optimized criteria are presented, which include the ones in [\href{https://link.aps.org/doi/10.1103/PhysRevLett.127.060504}{Phys. Rev. Lett. \textbf{127}, 060504 (2021)}] as special cases. Detailed example is given to show that our criteria can detect bound entanglement that can not be identified by positive partial transpose criterion, with the explicit measurement operators to experimentally measure the corresponding $\Lambda$-moments.
\end{abstract}

\maketitle

\section{Introduction}
Quantum entanglement is one of the most profound features of quantum mechanics \cite{PhysRev47777,PhysPhys195,hhhh}. It is the basic resource for many tasks in quantum information processing and quantum computation \cite{nielsen2002quantum}, such as quantum dense coding \cite{PhysRevLett.69.2881}, clock synchronization \cite{PRL852010}, quantum teleportation \cite{PhysRevLett.70.1895}, quantum secret sharing \cite{PhysRevA.59.1829} and quantum cryptography \cite{PhysRevLett.67.661}. One of the fundamental problems in the theory of quantum entanglement \cite{GUHNE20091,RevModPhys82277} is the detection of entanglement for generic quantum systems.

Many separability criteria have been proposed to detect the quantum entanglement. Let $H_X$ denote the Hilbert space with dimension $d_X$. The realignment criterion \cite{qip219,qic193} says that the realigned matrix $\mathcal{R}\left(\rho_{AB}\right)$ of any bipartite separable states $\rho_{AB}\in H_A\otimes H_B$ satisfies $\left\|\mathcal{R}\left(\rho_{AB}\right)\right\|_1\leqslant 1$, where $\|M\|_1$ denotes the trace norm of the matrix $M$ defined by $\|M\|_1=\mathrm{Tr}\sqrt{MM^{\dagger}}$. The positive partial transpose $\left(\mathrm{PPT}\right)$ criterion \cite{PhysRevLett.77.1413} says that for any bipartite separable state $\rho_{AB}$, one has $\rho_{AB}^{T_B}\geq 0$, where $\rho_{AB}^{T_B}$ is the partial transposed matrix with $T_B$ denoting the partial transposition with respect to the subsystem $B$. The state $\rho_{AB}$ is entangled
if the PPT criterion is violated.

Nevertheless, the PPT criterion is not operational experimentally in general, as the partial transpose is not a physical operation. Instead the experimentally measurable PT-moments have been introduced to quantify the correlations in many-body systems \cite{PhysRevLett109130502}. For any bipartite state $\rho_{AB}$, the PT-moments are given by $p_k=\mathrm{Tr}[(\rho_{AB}^{T_B})^k]$, $k=0,1,2,\cdots,n$, where $p_0$ is the number of nonzero eigenvalues of $\rho_{AB}^{T_B}$. It has been shown that the first three PT-moments can be used to give a simple yet powerful separability criterion of a bipartite state $\rho_{AB}$. If $\rho_{AB}$ is a PPT state \cite{PhysRevLett125200501}, then
\begin{eqnarray}\label{int1}
p_3\geq p^2_2.
\end{eqnarray}
These PT-moments can be efficiently measured by randomized measurements \cite{PhysRevLett125200501}.

Let $\bm{p}=(p_0,p_1,\cdots,p_d)$ be the vector given by all the PT-moments from $\rho_{AB}^{T_B}$, where $d=d_Ad_B$. A stronger criterion has been derived in \cite{PhysRevLett127060504} in terms of the higher order PT-moments. If $\rho_{AB}$ is a PPT state, then $B_l(\bm{p})\geq 0$ for $l=1,2,\cdots,\lfloor(d-1)/2\rfloor$. Here the Hankel matrix $B_{l}$ of order $l+1$ is defined with entries $[B_l(\bm{p})]_{ij}=p_{i+j+1}$ for $i,j=0,1,2,\cdots,l$ \cite{RAHCRJ,KS}. By reformulating the PT moment problem as an optimization, an explicit necessary condition was derived from a separable state for $p_3$, and further generalized to the cases of $p_k$ with $k>3$ \cite{PhysRevLett127060504}.

In fact, the PPT criterion is not a both necessary and sufficient condition of separability in general. A necessary and sufficient separability condition is given by positive but not completely positive maps $\Lambda$. Any state $\rho_{AB}$ is separable if and only if $(I_A\otimes\Lambda)(\rho_{AB})\geq0$ for all positive but not completely positive maps $\Lambda$ \cite{HORODECKI19961}, where $I_{A}$ is the identity map on $H_A$.

In this paper, we introduce $\Lambda$-moments given by the positive but not completely positive maps $\Lambda$ to study the detection of entanglement. We obtain general separability criteria based on $\Lambda$-moments, which give to the
separability criterion based on the PT-moments \cite{PhysRevLett125200501,PhysRevLett127060504} when $\Lambda$ is taken as the partial transpose. Our criteria are operational in the sense that they can be applied to certify the entanglement of unknown quantum states by physical measurements, as the $\Lambda$-moments can be estimated experimentally with a noiseless network \cite{PhysRevA74052323,PhysRevLett94040502,PhysRevA67060101}.
The paper is organized as follows. In Section \ref{seca} we introduce the $\Lambda$-moments. We present two kinds of entanglement criteria based on the $\Lambda$-moments. The first criterion is given by Hankel matrices, which is both necessary and sufficient for separability. The second criterion is a particularly optimized version from the first criterion. By using a detailed positive but not completely positive map, we show that our criterion can detect the entanglement of a bound entangled state better than the criterion from PT-moments. Moreover, the corresponding measurement operators are explicitly given to experimentally measure the $\Lambda$-moments for unknown quantum states. We conclude in the last section \ref{secc}.

\section{$\Lambda$-moments and entanglement detection}\label{seca}
A linear map $\Lambda$ acting on $H_B$ is called positive if it maps hermitian operators $X$ in $H_B$ onto hermitian ones such that $\Lambda(X^{\dagger})={\Lambda(X)}^{\dagger}$ and $\Lambda(X)\geq0$ for any $X\geq 0$, where $\dagger$ denotes the transpose and complex conjugation. A positive map $\Lambda$ is called completely positive if the map $I_A\otimes\Lambda$ acting on $H_A\otimes H_B$ is still positive. Otherwise, $\Lambda$ is said to be positive but not completely positive \cite{HORODECKI19961,GUHNE20091}.

Let $\Lambda$ be a linear hermiticity-preserving map on $H_B$.
For an arbitrary bipartite state $\rho_{AB}\in H_A\otimes H_B$, it is straightforward to prove that $I_A\otimes\Lambda(\rho_{AB})$ is Hermitian with real eigenvalues. Denote
\begin{equation}\label{lem11}
\Theta(\rho_{AB})=\frac{I_A\otimes\Lambda(\rho_{AB})}{\mathrm{Tr}[I_A\otimes\Lambda(\rho_{AB})]}.
\end{equation}
We define the $\Lambda$-moments associated with the positive map $\Lambda$ and state $\rho_{AB}$ as
\begin{equation}\label{def1}
q_k=\mathrm{Tr}[\Theta(\rho_{AB})^k],~~k=0,1,2,\cdots,n.
\end{equation}

Let the eigenvalue decomposition of $\Theta(\rho_{AB})$ be given by
$\Theta(\rho_{AB})=\sum_{i=1}^d\lambda_i|\lambda_i\rangle\langle\lambda_i|$.
Then from (\ref{def1}) we have $q_k=\sum_{i=1}^d\lambda_i^k$, $k=0,1,2,\cdots,d$.
Note here $q_0=\#\{\lambda_i\neq0\}$ is just the number of nonzero eigenvalues with $1\leq q_0\leq d$.
Moreover, we can see that $q_1=\sum_{i=1}^d\lambda_i=1$ for any positive map $\Lambda$.
As a special case, when $\Lambda=T_B$ is the partial transpose with respect to the subsystem $H_B$, the $\Lambda$-moments reduce to the PT-moments given in \cite{PhysRevLett125200501,PhysRevLett127060504}.

These $\Lambda$-moments $q_k$ can be experimentally measured for unknown quantum states. They can be expressed as the mean values of certain quantum mechanical observables \cite{PhysRevA74052323}. The mean values of the corresponding observables for an unknown state can be estimated with the help of positive but not completely positive maps by constructing noiseless networks \cite{PhysRevLett94040502}. The measurement is required only on the controlled qubit \cite{PhysRevA67060101}. This fact is in good agreement with the further proof that a single qubit may serve as interfaces connecting quantum devices \cite{PhysRevA69012305}. More details of the general scheme for constructing a noiseless network to estimate the mean value of observables with respect to an unknown state without its prior reconstruction can be found in \cite{PhysRevA74052323,PhysRevLett94040502,PhysRevA67060101}.

Denote $\bm{q}=(q_0, q_1,\cdots, q_d)$ the vector given by the $\Lambda$-moments defined in (\ref{def1}). The Hankel matrices $B_l(\bm{q})$ are $(l+1)\times(l+1)$ ($l\in\mathbb{N}_+$) matrices with the entries given by $[B_l(\bm{q})]_{ij}=q_{i+j+1}$ for $i,j=0,1,2,\cdots,l$.
The Hankel matrices $B_l(\bm{q})$ $(l\geq1)$ can be written as \cite{GHKR,EET},
\begin{equation}\label{01theorem}
B_l(\bm{q})=V_lDV_l^T,
\end{equation}
where
\begin{equation}\label{02theorem}
V_l=\begin{pmatrix}
 1 & 1 & \dots & 1 & \\
 \lambda_1 & \lambda_2 & \cdots & \lambda_d & \\
 \vdots & \vdots & \ddots & \vdots\\
 \lambda_1^l & \lambda_2^l & \cdots & \lambda_d^l &
\end{pmatrix},
\end{equation}
and $D=\mathrm{diag}\{\lambda_1,\lambda_2,\cdots,\lambda_d\}$.

$\mathit{Theorem\ 1}$.
A state $\rho_{AB}\in H_A\otimes H_B$ is separable if and only if $B_l(\bm{q})\geq0$ for $l\in\mathbb{N}_+$ and all positive maps $\Lambda$.

$\mathit{Proof}$.
If $\rho_{AB}$ is separable, then $\Theta(\rho_{AB})\geq0$ for any $\Lambda$, that is, $\lambda_i\geq0$ for $i=1,\cdots,d$ from the positive maps criterion \cite{HORODECKI19961}. Let $\bm{x}=(x_1,x_2,\cdots,x_{l+1})$ be an arbitrary vector belong to $\mathbb{R}^{l+1}$.
We have
\begin{equation}\label{03theorem}
\bm{x}B_l(\bm{q})\bm{x}^T=\bm{y}D\bm{y}^T=\sum_{i=1}^d\lambda_iy_i^2\geq0,
\end{equation}
where the first equality is due to (\ref{01theorem}), $\bm{y}=\bm{x}V_l=(y_1,y_2,\cdots,y_d)$ with $y_i=x_1+\sum_{j=2}^{l+1}x_j\lambda_i^{j-1}$, $i=1,\cdots,d$. Hence, If $\rho_{AB}$ is separable, then $B_l(\bm{q})\geq0$.

If $\rho_{AB}$ is an entangled state, there must exist a positive map $\Lambda$ such that $\Theta(\rho_{AB})$ is not positive \cite{HORODECKI19961}. Without loss of generality, suppose $\Theta(\rho_{AB})$ has $r$ different nonzero eigenvalues $\lambda_i$ with multiplicity $k_i$, $i=1,\cdots,r$, i.e., $\sum_{i=1}^rk_i=q_0$. At least one eigenvalue is negative.
Let $l'=r-1$. Then $B_{l'}(\bm{q})=V_{l'}D_{l'}V_{l'}^T$, where $V_{l'}$ is given by
\begin{equation}
V_{l'}=\begin{pmatrix}
 1 & 1 & \dots & 1 & \\
 \lambda_1 & \lambda_2 & \cdots & \lambda_r & \\
 \vdots & \vdots & \ddots & \vdots\\
 \lambda_1^{r-1} & \lambda_2^{r-1} & \cdots & \lambda_r^{r-1} &
\end{pmatrix}
\end{equation}
and $D_{l'}=\mathrm{diag}\{k_1\lambda_1,\cdots,k_r\lambda_r\}$. There must exist one vector $\bm{y}_1$ belong to $\mathbb{R}^r$ such that $\bm{y}_1D_{l'}\bm{y}_1^T<0$ since at least one
$k_i\lambda_i$, $i=1,\cdots,r$, is negative. According the property of the Vandermonde determinant, the Vandermonde matrix $V_{l'}$ is invertible. We have $\bm{x}_1B_{l'}(\bm{q})\bm{x}_1^T=\bm{x}_1V_{l'}D_{l'}V_{l'}^T\bm{x}_1^T =\bm{y}_1D_{l'}\bm{y}_1^T<0$, where $\bm{x}_1=\bm{y}_1V_{l'}^{-1}$. Therefore, $B_{l'}(\bm{q})\geq 0$ is not true. This is a contradiction. $\hfill\qedsymbol$

In particular, by noting that $q_1=1$ for any positive map $\Lambda$, the condition $B_1(\bm{q})\geq0$ from Theorem 1 gives rise to the following conclusion.

$\mathit{Corollary}$. If $\rho_{AB}$ is separable, then
\begin{equation}\label{corr}
q_3-q_2^2\geq0
\end{equation}
for any positive map $\Lambda$.

Obviously, the relation (\ref{int1}) is a special case of (\ref{corr}) with $\Lambda=T_B$. For brevity, we will refer to (\ref{corr}) as the $q_3$-$\Lambda$ criterion in the following.

From Theorem 1, there always exist positive maps such that the inequality $B_l(\bm{q})\geq0$ is violated for an entangled state. Nevertheless, the ability of detecting the entanglement of given states is different for different positive maps. Thus the separability problem is equivalent to finding the optimal positive maps with respect to the concerned entangled states \cite{stormer1963positive,WORONOWICZ1976165,CHOI1975285,maurer1977positive}. The classification of positive maps has been intensively studied from the perspective of quantum information theory. New positive but not completely positive maps have been derived, resulting in new separability criteria \cite{TERHAL200161,PhysRevLett97080501,chruscinski2007structure,PhysRevA73012345,CHRUSCINSKI20092301,hou2010characterization,Qi2011}.

The criteria given in Theorem 1 can be further optimized by adopting the method in \cite{PhysRevLett127060504}. The key idea to find the optimal criteria is based on the following optimization:
\begin{align}\label{opt0}
\min_{\lambda_i}/\max_{\lambda_i}\quad &\hat{q}_k:=\sum_{i=1}^d\lambda_i^k,\nonumber\\
\mbox{subject to}\quad
&\sum_{i=1}^d\lambda_i^n=q_n\ \mathrm{for}\ n=0,1,2,\cdots,k-1,\nonumber\\
&\lambda_i\geq0\ \mathrm{for}\ i=1,2,\cdots,d.
\end{align}
Once we obtain the optimal values $\hat{q}_k^{\min}$ and $\hat{q}_k^{\max}$, the relation $q_k\in[\hat{q}_k^{\min},\hat{q}_k^{\max}]$ provides a necessary condition for $\rho_{AB}$ to be separable.

For simplicity, we only consider the first three $\Lambda$-moments. The conditions $d\geq q_0\geq\lceil1/q_2\rceil$ and $\frac{1}{d}\leq q_2\leq1$ are always satisfied by all (separable or entangled) states. The condition $q_3\leq\hat{q}_3^{\max}$ is also redundancy since the maximization is still achieved without the constraints $\lambda_i\geq0$ for any states \cite{PhysRevLett127060504}. We denote
$\bm{\lambda}=(\lambda_1,\lambda_2,\cdots,\lambda_d)$ the vector given by the eigenvalues of $\Theta(\rho_{AB})$ in descending order, $\lambda_i\geq\lambda_{i+1}\geq0, i=1,\cdots,d-1$. Therefore, we simplify the optimization for the case $k=3$ as follows,
\begin{align}\label{opt1}
\min_{\lambda_i}\quad &\hat{q}_3:=\sum_{i=1}^d\lambda_i^3,\nonumber\\
\mbox{subject to}\quad
&\sum_{i=1}^d\lambda_i=q_1,\nonumber\\
&\sum_{i=1}^d\lambda_i^2=q_2,\nonumber\\
&\lambda_1\geq\lambda_2\geq\cdots\geq\lambda_d\geq0.
\end{align}
The minimization is attained at
\begin{equation}\label{opt3}
\bm{\lambda}_3^{\min}=(\lambda_1,\cdots,\lambda_1,\lambda_{\alpha+1},0,\cdots,0),
\end{equation}
where $\lambda_1$ appears $\alpha=\lfloor 1/q_2\rfloor$ times in (\ref{opt3}). The number of nonzero eigenvalues in (\ref{opt3}) is $\alpha+1$, i.e., $q_0=\alpha+1$. Then the minimized value is achieved at the minimum of $q_0$. Hence, we have the following theorem for every $\Lambda$-moments of order $3$.

$\mathit{Theorem\ 2}$.
If $\rho_{AB}\in H_A\otimes H_B$ is separable, then for any positive map $\Lambda$ the $\Lambda$-moment $q_3$ satisfies the following relation,
\begin{equation}\label{theorem31}
q_3\geq\alpha x^3+(1-\alpha x)^3,
\end{equation}
where $x=[\alpha+\sqrt{\alpha[(\alpha+1)q_2-1]}]/[\alpha(\alpha+1)]$ with $\alpha=\lfloor 1/q_2\rfloor$.

Theorem 2 includes the criteria in \cite{PhysRevLett127060504} as a special case of $\Lambda=T_B$. Obviously, (\ref{theorem31}) is also independent of the dimension $d$.
Comparing (\ref{corr}) with (\ref{theorem31}), we have $q_3\geq\alpha x^3+(1-\alpha x)^3\geq q_2^2$ for any separable states, in which the last equality holds when $\alpha=1/q_2$. Thus, we will refer to (\ref{theorem31}) as the $q_3$-O$\Lambda$ (optimal $\Lambda$-moments) criterion.
Analogous optimal criteria can be obtained for $\Lambda$-moments with $k=4,5,...,d$.
In what follows, we give a concrete example to illustrate the effectiveness of our criteria.
We also present the corresponding measurement operators, as well as specific observables which are experimentally feasible by constructing noiseless networks \cite{PhysRevA74052323,PhysRevLett94040502,PhysRevA67060101}.

$\mathit{Example}$. Consider the Horodecki's $3\times3$ bound entangled state,
\begin{equation}\label{exp1}
\sigma_{a}=\frac{2}{7}\ket{\Psi^{+}} \la\Psi^+|+\frac{a}{7}\sigma_{+}+\frac{5-a}{7}\sigma_{-},
\end{equation}
where
\begin{align}
\ket{\Psi^+}&=\frac{1}{\sqrt{3}}(\ket{00}+\ket{11}+\ket{22}),\nonumber\\
\sigma_+&=\frac{1}{3}(|01\rangle\langle01|+|12\rangle\langle12|+|20\rangle\langle20|),\nonumber\\
\sigma_-&=\frac{1}{3}(|10\rangle\langle10|+|21\rangle\langle21|+|02\rangle\langle02|).\nonumber
\end{align}
The state is separable for $2\leq a\leq3$, bound entangled for $3<a\leq4$ and free entangled for $4<a\leq5$ \cite{PhysRevLett821056}.

Let us consider a concrete positive map $\Lambda_1$ that maps any $3\times 3$ matrix $A$ with entries $(A)_{ij}=a_{ij}$ to a matrix with entries $[\Lambda_1(A)]_{ij}=-a_{ij}$ for $i\neq j$, and $[\Lambda_1(A)]_{ii}=a_{ii}+a_{i'i'}$ for $i=j$, where $i'=i+2\bmod 3$. $\Lambda_1$ is a positive but not completely positive map \cite{hou2010characterization,Qi2011}.
In this case, we have $\mathrm{Tr}[I_A\otimes\Lambda_1(\sigma_a)]=2$ and
\begin{equation}\label{exp2}
\Theta(\sigma_a)=\frac{1}{2}I_A\otimes\Lambda_1(\sigma_a)
\end{equation}
from (\ref{lem11}). From (\ref{def1}) we have $q_k=\mathrm{Tr}[\Theta(\sigma_a)^k]$
with $k=0,1,2,\cdots,n$.

\begin{figure}[t]
  \centering
  \includegraphics[width=8cm]{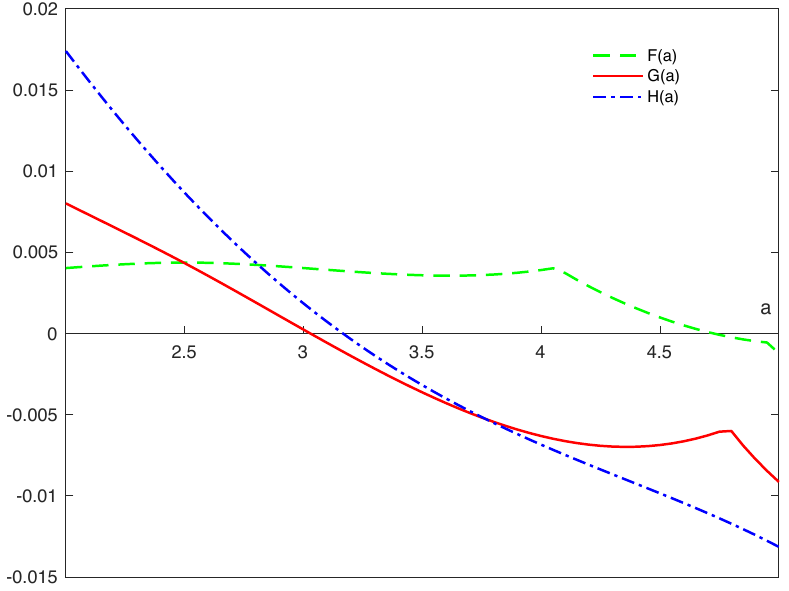}\\
  \caption{Entanglement detection based on $\Lambda_1$-moments and PT-moments. The dot dashed (blue) line is for the $q_3$-$\Lambda_1$ criterion, and the solid (red) line is for the $q_3$-O$\Lambda_1$ criterion. The dashed (green) line is for the $p_3$-OPPT criterion.}
  \label{f1}
\end{figure}

If $\sigma_{a}$ is separable, we have $H(a)\equiv q_3-q_2^2\geq0$ from the $q_3$-$\Lambda_1$ criterion defined in (\ref{corr}), $G(a)\equiv q_3-(\alpha x^3+(1-\alpha x)^3)\geq 0$ from the $q_3$-O$\Lambda_1$ criterion defined in (\ref{theorem31}), where $\alpha=\lfloor 1/q_2\rfloor$ and $x=[\alpha+\sqrt{\alpha[(\alpha+1)q_2-1]}]/\alpha(\alpha+1)$, and $F(a)\equiv p_3-(\beta y^3+(1-\beta y)^3)\geq 0$ from the $p_3$-OPPT criterion based on the PT-moments \cite{PhysRevLett125200501,PhysRevLett127060504},
where $p_k=\mathrm{Tr}[(\sigma_a^{T_B})^k]$, $k=0,1,2,\cdots,n$, $\beta=\lfloor 1/p_2\rfloor$ and $y=[\beta+\sqrt{\beta[(\beta+1)p_2-1]}]/[\beta(\beta+1)]$.
Fig. \ref{f1} shows that the $q_3$-$\Lambda_1$ criterion detects entanglement for $a\in[3.1658, 5]$. The $q_3$-O$\Lambda_1$ criterion detects entanglement for $a\in[3.0291, 5]$. While the $p_3$-OPPT criterion only detects entanglement for $a\in[4.7259, 5]$.

Obviously, the $q_3$-$\Lambda_1$ is slightly weaker than the $q_3$-O$\Lambda_1$, but still stronger than the $p_3$-OPPT criterion for the states (\ref{exp1}).

Here, all the $\Lambda_1$-moments can experimentally measured. Hence, in verifying the entanglement of an unknown quantum state, one does not need to take tomography. By using $k$ copies of the state $\sigma_{a}$, the $\Lambda_1$-moments $q_k$ in the above example can be expressed as $q_k=\mathrm{Tr}[\mathcal{O}^{(k)}\sigma_a^{\otimes k}]$, where
$\mathcal{O}^{(k)}=(V^{(k)}+V^{(k)\dag})/2^{k+1}$ with $V^{(k)}=V_A^{(k)}\otimes V_B^{(k)}$.
$V_A^{(k)}=\overrightarrow{\Pi}_A$ and $V_B^{(k)}=(\Lambda_1^\dag)^{\otimes k}(\overrightarrow{\Pi}_B)$ with $\overrightarrow{\Pi}_A$ and $\overrightarrow{\Pi}_B$ the $k$ copies cyclic permutation operators, i.e., $\overrightarrow{\Pi}\ket{l^1, l^2,\cdots, l^k}=\ket{l^k, l^1,\cdots, l^{k-1}}$. The measurement can be carried out as long as the expression of $V_B^{(k)}$ is obtained. In particular, with respect to $q_2$ and $q_3$ in the example, we have $V_B^{(2)}=I_B+\overrightarrow{\Pi}_B$ and $V_B^{(3)}=O_1+O_2+O_3$,
where
\begin{align}
O_1=&\sum_j\{|jjj\rangle\langle jjj|+|jjj'\rangle\langle jjj'|+|jj'j\rangle\langle jj'j|\nonumber\\
&+|jj'j'\rangle\langle jj'j'|+|j'jj\rangle\langle j'jj|+|j'jj'\rangle\langle j'jj'|\nonumber\\
&+|j'j'j\rangle\langle j'j'j|+|j'j'j'\rangle\langle j'j'j'|\},\nonumber
\end{align}
\begin{align}
O_2=&\sum_{j\neq l}\{|jjl\rangle\langle jlj|+|j'jl\rangle\langle j'lj|+|jll\rangle\langle llj|\nonumber\\
&+|jl'l\rangle\langle ll'j|+|jlj\rangle\langle ljj|+|jlj'\rangle\langle ljj'|\},\nonumber
\end{align}
\begin{align}
O_3=-\sum_{j\neq l\neq v}|jlv\rangle\langle lvj|,\nonumber
\end{align}
with $j,l,v=0,1,2$, $j'=j+2\bmod 3$ and $l'=l+2\bmod 3$.

\section{Conclusions and discussions}\label{secc}
We have introduced the $\Lambda$-moments associated any positive map $\Lambda$ to effectively characterize the entanglement of unknown quantum states without theirs prior reconstructions.
Based on these $\Lambda$-moments we have presented necessary and sufficient conditions for separability of quantum states. The criteria based on PT-moments, which are not both necessary and sufficient, are special cases of our results. In addition, we have given the optimized criterion related to the $\Lambda$-moment $q_3$ by adopting the method used in \cite{PhysRevLett127060504}. Our approach may also highlight the characterization of multipartite entanglement, although it remains open to evaluate the corresponding moments by using randomized measurements.

\bigskip
\section*{Acknowledgments} This work is supported by the National Natural Science Foundation of China (NSFC) under Grants 12075159 and 12171044; Beijing Natural Science Foundation (Grant No. Z190005); the Academician Innovation Platform of Hainan Province; Shenzhen Institute for Quantum Science and Engineering, Southern University of Science and Technology (No. SIQSE202001).

\smallskip
\section*{Data Availability Statement}
All data generated or analysed during this study are included in this published article (and its supplementary information files).


\smallskip
\begin{thebibliography}{99}
\bibitem{PhysRev47777}A. Einstein, B. Podolsky, N. Rosen, \href{https://link.aps.org/doi/10.1103/PhysRev.47.777}{Phys. Rev. \textbf{47}, 777 (1935)}

\bibitem{PhysPhys195}J. S. Bell, \href{https://link.aps.org/doi/10.1103/PhysPhys.195}{Phys. Phys. Fiz. \textbf{1}, 195 (1964)}

\bibitem{hhhh} R. Horodecki, P. Horodecki, M. Horodecki, K. Horodecki, \href{https://link.aps.org/doi/10.1103/RevModPhys.81.865}{Rev. Mod. Phys. \textbf{81}, 865 (2009)}

\bibitem{nielsen2002quantum}M. A. Nielsen, I. Chuang, Quantum Computation and Quantum Information (2002)

\bibitem{PhysRevLett.69.2881}C. H. Bennett, S. J. Wiesner, \href{https://link.aps.org/doi/10.1103/PhysRevLett.69.2881}{Phys. Rev. Lett. \textbf{69}, 2881 (1992)}

\bibitem{PRL852010}R. Jozsa, D. S. Abrams, J. P. Dowling, C. P. Williams, \href{https://link.aps.org/doi/10.1103/PhysRevLett.85.2010}{Phys. Rev. Lett. \textbf{85}, 2010 (2000)}

\bibitem{PhysRevLett.70.1895}C. H. Bennett, G. Brassard, C. Cr\'epeau, R. Jozsa, A. Peres, W. K. Wootters, \href{https://link.aps.org/doi/10.1103/PhysRevLett.70.1895}{Phys. Rev. Lett. \textbf{70}, 1895 (1993)}

\bibitem{PhysRevA.59.1829}M. Hillery, V. Bu\ifmmode \check{z}\else \v{z}\fi{}ek, A. Berthiaume, \href{https://link.aps.org/doi/10.1103/PhysRevA.59.1829}{Phys. Rev. A \textbf{59}, 1829 (1999)}

\bibitem{PhysRevLett.67.661}A. K. Ekert, \href{https://link.aps.org/doi/10.1103/PhysRevLett.67.661}{Phys. Rev. Lett. \textbf{67}, 661 (1991)}

\bibitem{GUHNE20091}O. G{\"u}hne, G. T{\'o}th, \href{https://www.sciencedirect.com/science/article/pii/S0370157309000623}{Phys. Rep. \textbf{474}, 1 (2009)}

\bibitem{RevModPhys82277}J. Eisert, M. Cramer, M. B. Plenio, \href{https://link.aps.org/doi/10.1103/RevModPhys.82.277}
{Rev. Mod. Phys. \textbf{82}, 277 (2010)}


\bibitem{qip219} O. Rudolph, \href{https://doi.org/10.1007/s11128-005-5664-1}{Quantum Inf. Process. \textbf{4} 219 (2005)}

\bibitem{qic193} K. Chen, L.A. Wu, \href{https://doi.org/10.1103/PhysRevE.67.056305}{Quantum Inf. Comput. \textbf{3}, 193 (2003)}

\bibitem{PhysRevLett.77.1413} A. Peres, \href{https://doi.org/10.1103/PhysRevLett.77.1413}{Phys. Rev. Lett. \textbf{77}, 1413 (1996)}

\bibitem{PhysRevLett109130502} P. Calabrese, J. Cardy, E. Tonni, \href{https://link.aps.org/doi/10.1103/PhysRevLett.109.130502}{Phy. Rev. Lett. \textbf{109}, 130502 (2012)}

\bibitem{PhysRevLett125200501} A. Elben, R. Kueng, H.Y.R. Huang, R. van Bijnen, C. Kokail, M. Dalmonte, P. Calabrese, B. Kraus, J. Preskill, P. Zoller, B. Vermersch, \href{https://link.aps.org/doi/10.1103/PhysRevLett.125.200501}{Phys. Rev. Lett.\textbf{125}, 200501 (2020)}

\bibitem{PhysRevLett127060504} X. D. Yu,  S. Imai, O. G{\"u}hne, \href{https://link.aps.org/doi/10.1103/PhysRevLett.127.060504}{Phys. Rev. Lett. \textbf{127}, 060504 (2021)}

\bibitem{RAHCRJ}R. A. Horn, C. R. Johnson, Matrix analysis (Cambridge University Press, 2012)

\bibitem{KS}K. Schmüdgen, The moment problem (Springer, 2017)

\bibitem{HORODECKI19961} M. Horodecki, P. Horodecki, R. Horodecki, \href{https://doi.org/10.1016/S0375-9601(96)00706-2}{Phys. Lett. A \textbf{223}, 1 (1996)}

\bibitem{PhysRevA74052323} P. Horodecki, R. Augusiak, M. Demianowicz, \href{https://link.aps.org/doi/10.1103/PhysRevA.74.052323}{Phys. Rev. A \textbf{74}, 052323 (2006)}

\bibitem{PhysRevLett94040502} H. A. Carteret, \href{https://link.aps.org/doi/10.1103/PhysRevLett.94.040502}{Phys. Rev. Lett. \textbf{94}, 040502 (2005)}

\bibitem{PhysRevA67060101} P. Horodecki, \href{https://link.aps.org/doi/10.1103/PhysRevA.67.060101}{Phys. Rev. A \textbf{67}, 060101(R) (2003)}

\bibitem{PhysRevA69012305}S. Lloyd, A. J. Landahl, J.-J. E. Slotine, \href{https://link.aps.org/doi/10.1103/PhysRevA.69.012305}
{Phys. Rev. A \textbf{69}, 012305 (2004)}

\bibitem{GHKR}G. Heinig, K. Rost, In Algebraic Methods for Toeplitz-like Matrices and Operators (De Gruyter, 2022)

\bibitem{EET}E. E. Tyrtyshnikov, \href{https://doi.org/10.1007/s002110050027}{Numerische Mathematik \textbf{67}, 261 (1994)}

\bibitem{stormer1963positive}E. St{\o}rmer, Acta Math. \textbf{110}, 233 (1963)

\bibitem{WORONOWICZ1976165}S. Woronowicz, \href{https://doi.org/10.1016/0034-4877(76)90038-0}{Rep. Math. Phys. \textbf{10}, 165 (1976)}

\bibitem{CHOI1975285}M.-D. Choi, \href{https://doi.org/10.1016/0024-3795(75)90075-0}{Linear Algebra Appl. \textbf{10}, 285 (1975)}

\bibitem{maurer1977positive}J. Maurer, \href{https://doi.org/10.1007/BF01223959}{Arch. Math. \textbf{28}, 510 (1977)}

\bibitem{hou2010characterization} J. Hou, \href{https://doi.org/10.1088/1751-8113/43/38/385201}{J. Phys. A \textbf{43}, 385201 (2010)}

\bibitem{TERHAL200161}B. M. Terhal, \href{https://doi.org/10.1016/S0024-3795(00)00251-2}{Linear Algebra Appl. \textbf{323}, 61 (2001)}

\bibitem{PhysRevLett97080501}H.-P. Breuer, \href{10.1103/PhysRevLett.97.080501}{Phys. Rev. Lett. \textbf{97}, 080501 (2006)}

\bibitem{chruscinski2007structure}D. Chru{\'s}ci{\'n}ski, A. Kossakowski, \href{https://doi.org/10.1007/s11080-007-9052-4}{Open Syst. Inf. Dyn. \textbf{14}, 275 (2007)}

\bibitem{PhysRevA73012345}M. Piani, \href{10.1103/PhysRevA.73.012345}{Phys. Rev. A \textbf{73}, 012345 (2006)}

\bibitem{CHRUSCINSKI20092301}D. Chru{\'s}ci{\'n}ski, A. Kossakowski, \href{https://doi.org/10.1016/j.physleta.2009.04.068}{Phys. Lett. A \textbf{373}, 2301 (2009)}

\bibitem{Qi2011}X. F. Qi, J. C. Hou, \href{https://doi.org/10.48550/arXiv.1008.3682}{J. Phys. A \textbf{44}, 215305 (2011)}

\bibitem{PhysRevLett821056} P. Horodecki, M. Horodecki, R. Horodecki, \href{https://link.aps.org/doi/10.1103/PhysRevLett.82.1056}{Phys. Rev. Lett. \textbf{82}, 1056 (1999)}


\end{thebibliography}
\end{document}